\documentclass[10pt]{iopart} 
\usepackage[utf8]{inputenc}
\usepackage{graphicx}
\usepackage[colorlinks,citecolor=blue,urlcolor=blue,linkcolor=blue]{hyperref}
\usepackage{siunitx}

\expandafter\let\csname equation*\endcsname\relax
\expandafter\let\csname endequation*\endcsname\relax

\usepackage{amsmath}
\usepackage{iopams}
\usepackage{widetable}
\usepackage{bm}
\usepackage{cite}
\usepackage{dsfont}
\usepackage{booktabs}
\usepackage{supertabular}
\usepackage{color}
\usepackage{hyperref}
\usepackage{mathrsfs}
\usepackage{float}
\usepackage{textcomp}
\usepackage{epsfig}
\usepackage{epstopdf}
\usepackage{multirow}
\usepackage{siunitx}
\usepackage{braket}
\usepackage[titletoc]{appendix}
\usepackage{colortbl,hhline}
\usepackage{graphicx}% Include figure files
\usepackage{mathtools}
\usepackage{subfigure}
\usepackage{dcolumn}% Align table columns on decimal point
\usepackage{bm}% bold math
\usepackage[markup=underlined]{changes}
\definechangesauthor[color=orange]{TG}

\newcommand\newblock{\hskip .11em\@plus.33em\@minus.07em}

\begin{document}

\title[]{Optimizing the spectro-temporal properties of photon pairs from Bragg-reflection waveguides}
\author{H. Chen$^{1,2}$,  K. Laiho$^3$, B. Pressl$^{2,4}$, A. Schlager$^2$, \\ H. Suchomel$^{5}$, M. Kamp$^{5}$, S. H\"ofling$^{5,6}$, C. Schneider$^{5}$, \\ and G. Weihs$^2$}
\address{$^1$Department of Physics, National University of Defense Technology, Changsha, 410073, People's Republic of China}
\address{$^2$Institut f\"ur Experimentalphysik, Universit\"at Innsbruck, Technikerstra\ss e 25, 6020 Innsbruck, Austria}
\address{$^3$Technische Universit\"at Berlin, Institut f\"ur Festk\"orperphysik, Hardenbergstr.~36, 10623 Berlin, Germany}
\address{$^4$Sektion f\"ur Biomedizinische Physik, Medizinische Universit\"at Innsbruck, M\"ullerstraße 44, 6020 Innsbruck, Austria}
\address{$^5$Technische Physik, Universit\"at W\"urzburg, Am Hubland,  97074 W\"urzburg, Germany}
\address{$^6$School of Physics $\&$ Astronomy, University of St Andrews, St Andrews, KY16 9SS, United~Kingdom}

%----------------------------------------------------------------------------------
%                            Abstract
%----------------------------------------------------------------------------------
\date{\today}

\begin{abstract}
Bragg-reflection waveguides (BRWs) fabricated from AlGaAs provide an interesting non-linear optical platform for photon-pair generation via parametric down-conversion (PDC). In contrast to many conventional PDC sources, BRWs are made of high refractive index materials and their characteristics are very sensitive to the underlying layer structure. First, we show that the design parameters like the phasematching wavelength and the group refractive indices of the interacting modes can be reliably controlled even in the presence of fabrication tolerances. We then investigate, how these characteristics can be taken advantage of when designing quantum photonic applications with BRWs. We especially concentrate on achieving a small differential group delay between the generated photons of a pair and then explore the performance of our design when realizing a Hong-Ou-Mandel interference experiment or generating spectrally multi-band polarization entangled states. Our results show that the versatility provided by engineering the dispersion in BRWs 
is important for employing them in different quantum optics tasks.
\end{abstract}

\ioptwocol

%----------------------------------------------------------------------------------
\section{Introduction}
%----------------------------------------------------------------------------------
Parametric down-conversion (PDC) offers means for comparably simple generation of non-classical light that can be used  as a robust information carrier in a variety of quantum optics applications \cite{caspani2017integrated, Flamini2018}. Nevertheless, the spectral properties of the  PDC emission, which largely stem from the material dispersion, play a crucial role in deciding for which applications a photon-pair source is suitable \cite{C.K.Hong1987, Evans2010, Zhu2012,Bruno2014a, Kang2014}. The intrinsic spectral characteristics of PDC  can even exclude the source from being directly applicable in a desired task, and therefore, spectral filtering is often employed to counter this drawback \cite{Laiho2011a, Bruno2014,Silverstone2014}. Otherwise, sophisticated shaping of the spectral extent of the PDC photons  based on modulation or non-linear effects can be employed \cite{Donohue2016, Wright2017, Puigibert2017, Ansari2018}. Still, the most direct and efficient approach is the modification of the source's intrinsic properties. This is also the case for state engineering in semiconductor Bragg-reflection waveguides (BRWs)\cite{Zhukovsky2012,Horn-scientific-reports-2013, kang2016monolithic, Claire2016Multi-user,schlager--2017--temporally}.

BRWs are usually made of AlGaAs, which is an excellent integrated optics platform and possesses a large second-order optical nonlinearity. The photon pairs---signal and idler---emitted by PDC in BRWs, which need to fulfill energy and momentum conservation, can be engineered to be counter- \cite{lanco2006semiconductor} or co-propagating \cite{sarrafi2013continuous, Gregor-Monolithic-Source-2012}. We are interested in the latter case, in which both the fundamental and higher order spatial modes propagating in the structure are utilized \cite{West--2006--Analysis-BRW,Helmy--2006--Phase-matching}. However, adapting BRW sources to specific quantum optics tasks requires suitable joint spectral properties of signal and idler. For this purpose, one can benefit from the lack of birefringence in the underlying semiconductor platform. In addition, the strong dispersion of these high refractive index materials \cite{Gehrsitz-refractive-index-2000} can be taken advantage of. Luckily, accurate experimental methods exist for the verification of the group refractive indices and their dispersion in these multimode waveguides \cite{Pressl--2015,laiho--2016--uncovering, Misiaszek--2018}.

In this paper, we perform a detailed simulation of the optical properties of the PDC emission from the BRW sample designed in Ref.~\cite{Pressl-2018-Semi-automatic}. We investigate its performance in specific quantum optics applications that rely on  spectral and temporal indistinguishability of the PDC photons. We start in Section \uppercase\expandafter{2} by exploring the robustness of our design with respect to variations in its structure. In Section \uppercase\expandafter{3}, we  explore, how to minimize the average differential group delay between signal and idler. In Section \uppercase\expandafter{4}, we show that the PDC emission from our BRW design produces a large spectral overlap between the photon pairs. Thereafter,  in Section \uppercase\expandafter{5} we study the performance of our PDC source in two quantum optics tasks. We show that due to the optimized spectral and temporal overlap of signal and idler, our design can be used as a versatile source in Hong-Ou-Mandel (HOM) interference experiments and that it is well suited for the generation of spectrally multi-band polarization entanglement.

%----------------------------------------------------------------------------------
\section{Effect of variations in the BRW layer parameters on the phasematching wavelength}
%----------------------------------------------------------------------------------
\begin{figure}
\center
\includegraphics[width = 0.45\textwidth]{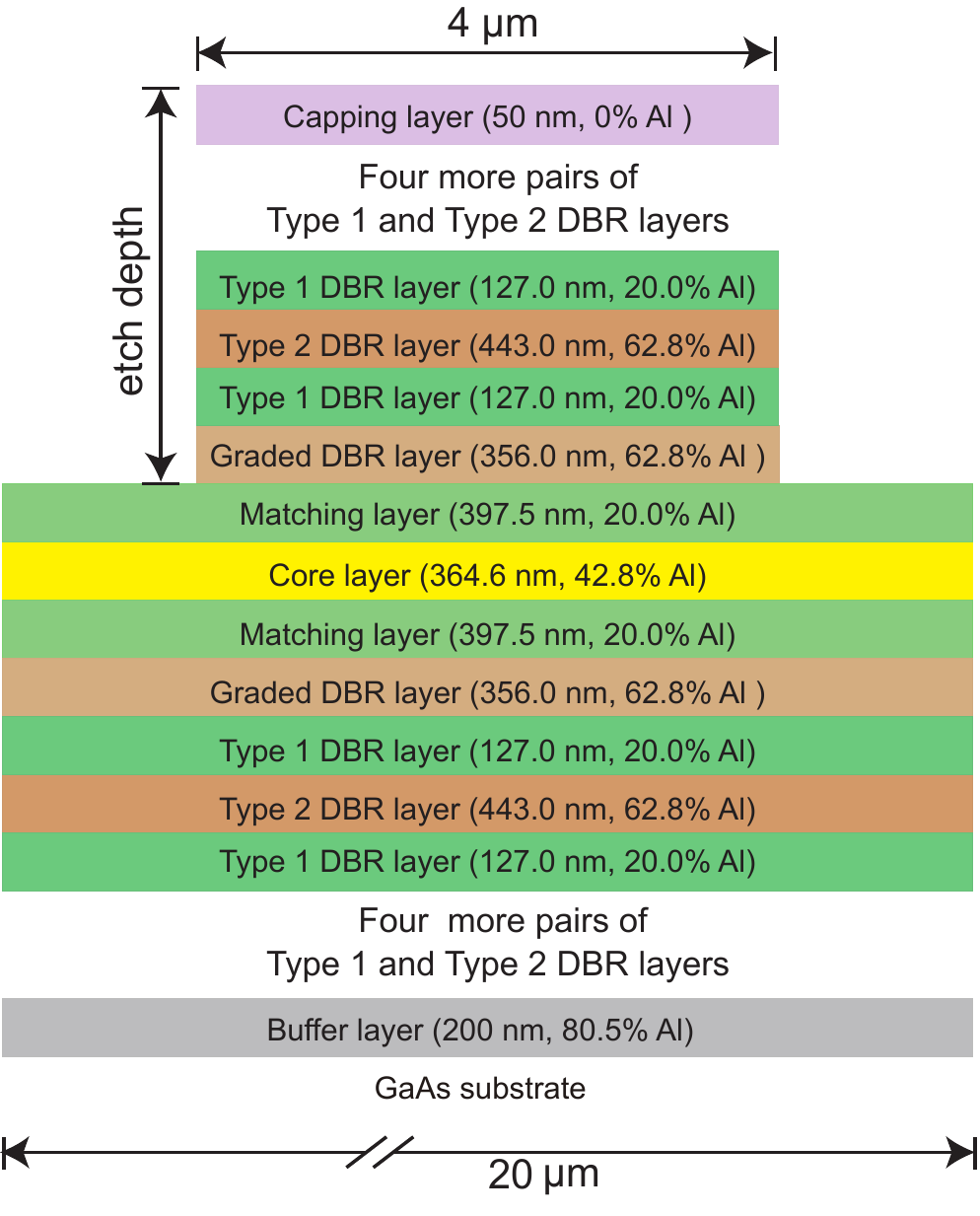}
\caption{(Color online): Layer structure of the investigated Bragg-reflection waveguide with different kinds of layers visualized with different colors. The waveguide core (yellow slab) is enclosed by the matching layers. The distributed Bragg-reflector (DBR) stacks surround this region on top and bottom. The DBR layers closest to the matching layers  are thinner and therefore called graded DBRs.
Each stack includes six(five)  type 1(2) DBR layers. The layer thicknesses and their aluminum contents are indicated in brackets.
\label{fig:new_structure}}
\end{figure}

First, we investigate the effect of certain geometric and material parameters on the operating wavelength of our sample. We  regard the \textit{graded} BRW design from Ref.~\cite{Pressl-2018-Semi-automatic}, which was optimized for simpler fabrication while maintaining a high efficiency for the desired PDC process. \autoref{fig:new_structure} shows the cross-section and main parameters of the structure. This ridge waveguide consists of a core layer (CL) and two matching layers (ML) around the core, which are surrounded by six pairs of distributed Bragg reflectors (DBRs). The phasematching condition is fulfilled by guiding the pump light in the so-called Bragg mode, which is a higher-order spatial mode, whereas the PDC photon pairs are generated in total internal reflection modes. The phasematching is defined by the dispersion of the modes and is heavily affected by the material composition and geometric layout of the ridge. Therefore, it is very sensitive to the structural parameters, i.e. the layer thicknesses and their aluminum contents as well as the width and height of the ridge. Such deviations may arise for example due to imperfections in the fabrication process.

We employ a commercial mode solver \cite{Comsol} to determine the dispersion of the eigenmodes. The nominal design, which is depicted in \autoref{fig:new_structure}, has a ridge width of \SI{4}{\micro\meter} and a height of approximately \SI{3.3}{\micro\meter} denoted here as the etch depth. We use the model of Gehrsitz et al.~in Ref.~\cite{Gehrsitz-refractive-index-2000} to estimate the refractive indices of the individual layers, which are greatly dependent on their aluminum concentrations. The graded BRW supports a type-II PDC process such that signal and idler are cross-polarized. After solving the dispersion of the interacting modes, the phasematched wavelength triplets are  found by searching for solutions of $n_s(\lambda_{s})/\lambda_{s} + n_i(\lambda_{i})/\lambda_{i} = n_p(\lambda_p)/\lambda_p$ with  $1/\lambda_{s} + 1/\lambda_{i} = 1/\lambda_{p}$, in which $n_{\mu}(\lambda_{\mu})$ ($\mu= s,i,p$) describe the effective refractive indices of the signal ($s$), idler ($i$) and pump ($p$) modes  in terms of the wavelength $\lambda$. At the degeneracy $\lambda_{s} = \lambda_{i} = 2\lambda_{p}$, which we call the phasematching wavelength. For the graded BRW as shown in \autoref{fig:new_structure} we find a phasematching wavelength of \SI{1553.8}{\nano\meter}.

\begin{table}
\begin{tabular}{lcc}
\br
  Parameter & (i) & (ii)  \\ \mr
  ML thickness &$+\SI{3.066}{\nano\meter}$  &  $+\SI{14.36}{\nano\meter}$\\ 
  ML Al content & $\SI{-0.287}{\nano\meter} $ & $\SI{-2.45}{\nano\meter}$ \\
  CL thickness & $+\SI{1.856}{\nano\meter} $ & $+\SI{9.30}{\nano\meter}$\\ 
  CL Al content & $\SI{-0.661}{\nano\meter} $ & $\SI{-3.46}{\nano\meter}$ \\ 
  Type 1 DBR thickness & $+\SI{0.098}{\nano\meter} $& $+\SI{0.53}{\nano\meter}$  \\ 
  Type 1 DBR Al content & $\SI{-0.244}{\nano\meter} $& $\SI{-0.61}{\nano\meter}$ \\
  Type 2 DBR thickness & $+\SI{0.044}{\nano\meter} $& $+\SI{0.22}{\nano\meter}$ \\
  Type 2 DBR Al content & $\SI{-0.027}{\nano\meter} $& $\SI{-0.27}{\nano\meter}$ \\ 
 %&\\
\br
\end{tabular}
\caption{\label{percent-change} Shift in the phasematching wavelength (i) caused by +\SI{1}{\percent} relative variation in the listed layer parameter and (ii) expected for the graded-BRW from  \autoref{fig:new_structure} within the fabrication tolerances ($+\SI{5}{\percent}$ deviation in the layer thickness or $+\SI{2}{}$ percentage point change in their Al contents). We note that the refractive index model is the main source of numerical imperfections in our approach. Due to the accuracy of $10^{-4}$ in the refractive indices, we report the expected phasematching wavelength shifts in (ii) with the precision of \SI{0.01}{\nano\meter}.}
\end{table}

We then vary each layer parameter listed in \autoref{percent-change} from its specified value  by $+$\SI{1}{\percent}, while keeping the others fixed, and calculate the change in the phasematching wavelength of the graded BRW  shown as case (i).  It is apparent, that an increment in the layer thicknesses also increases the phasematching wavelength, while when regarding the aluminum contents the opposite is observed. This effectively means that one can simply scale the structure to achieve another phasematching wavelength without having to re-engineer the epitaxial structure. Moreover, we remark that these sensitivities have to be considered in the context of the practically achievable fabrication accuracies. We expect thickness accuracies better than \SI{5}{\percent} and aluminum content accuracies of $2$ in absolute percentage points. As case (ii) in  \autoref{percent-change} we show the variation in the phasematching wavelength for the design in \autoref{fig:new_structure} caused by the fabrication tolerance in the listed layer parameter, while the others are kept constant.

\begin{figure}
  \centering
  \includegraphics[width=0.4\textwidth]{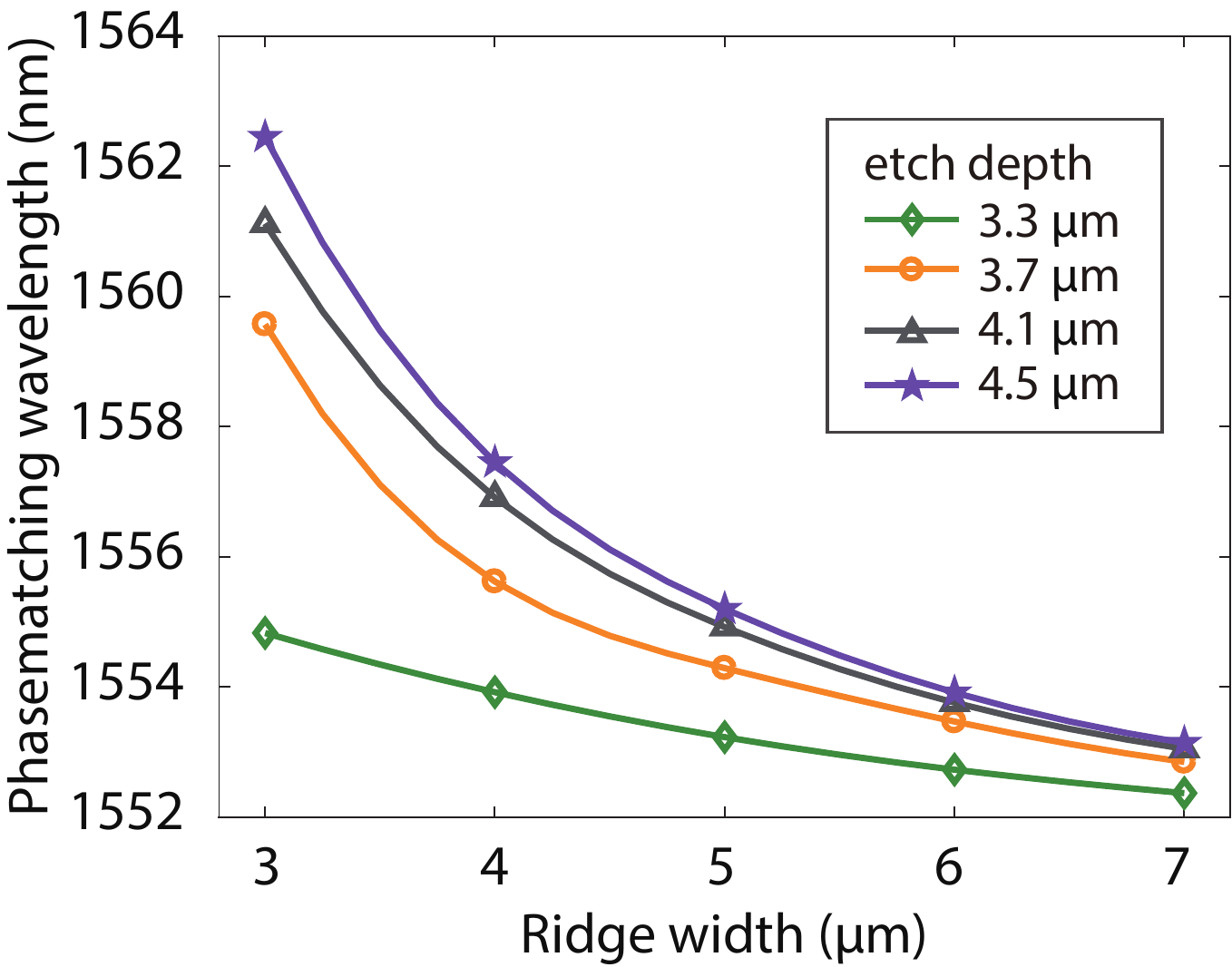}
  \caption{Phasematching wavelength in terms of the ridge width for different etch depths. The symbols illustrate the simulated values, while the solid lines provide guides for the eyes.}
  \label{etch-depth}
\end{figure}

The  ridge width \cite{Bijlani2008, Abolghasem2009} and etch depth allow some posterior changes of the phasematching wavelength. After the epitaxial growth of the layers, their thickness can be determined and the ridge width and height can be adjusted accordingly.  We show this effect in \autoref{etch-depth}, which illustrates the phasematching wavelength in terms of a typical range of ridge widths for different etch depths. The deeper etches cause the phasematching wavelength to depend stronger on the ridge width. Additionally, one can clearly see the increased effect of the stronger horizontal confinement of the modes for very narrow waveguides. Thus, the tuning range provided by employing different ridge widths greatly depends on the used etch depth.

%----------------------------------------------------------------------------------
\section{Temporal overlap of signal and idler}
%----------------------------------------------------------------------------------

While the effective refractive indices of signal, idler and pump determine the exact phasematching wavelength, the effective \textit{group refractive indices} of these modes play a crucial role when regarding the spectro-temporal properties of the created photon pairs. For example, if signal and idler group velocities differ from each other, their wavepackets walk temporally off during the propagation in the BRW, which  results in temporal distinguishability. Some applications such as the generation of polarization entanglement, which we investigate in more detail in Section  \uppercase\expandafter{5}, are very sensitive to this average differential group delay (DGD). Only if it is negligible, such a scheme can be implemented without external optical delay compensation \cite{schlager--2017--temporally, Chen2017}.
The average DGD between the signal and idler wavepackets is given by
\begin{equation}
 \textrm{DGD} = \frac{L}{2}\left|\frac{1}{\tilde{v}_s} - \frac{1}{\tilde{v}_i}\right| = \frac{L}{2c}\left| \tilde{n}_s- \tilde{n}_i\right|,
\label{eq:delta_t}
\end{equation}
 where $L$ is the length of the waveguide, $\tilde{v}_s$ and $\tilde{v}_i$ ($\tilde{n}_s$ and $\tilde{n}_i$) indicate the group velocities (group refractive indices) of signal and idler, respectively, and $c$ is the speed of light in vacuum. The group refractive index of each mode is defined as
\begin{equation}
\begin{split}
\tilde{n}_{\mu} &= n_{\mu} + \omega\frac{\mathrm{d} n_{\mu}}{ \mathrm{d}  \omega} 
\end{split}
\end{equation}
in terms of its effective refractive index and angular frequency $\omega = 2\pi c/\lambda$.

\begin{figure}
  \centering
  \includegraphics[width=0.42\textwidth]{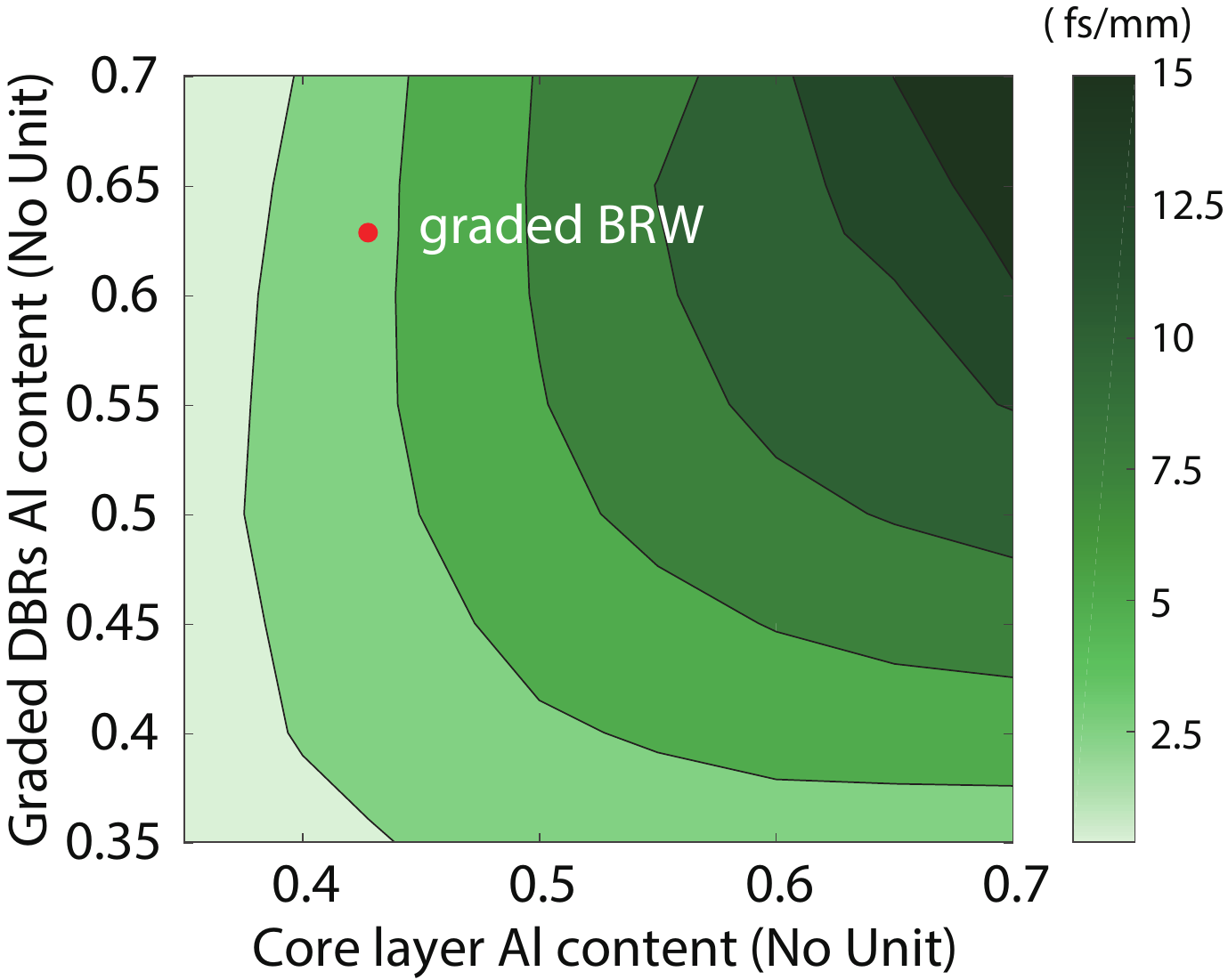}
  \caption{Simulated differential group delay (DGD) per length near \SI{1550}{\nano\meter} in terms of the aluminum content in the layers surrounding the matching layers. The red dot marks the position of the graded BRW structure  from \autoref{fig:new_structure} in terms of its respective Al contents.}
  \label{effective_index_difference}
\end{figure}

One usually resorts to numerical simulations in order to accurately predict the BRW properties since an analytical solution yielding the effective refractive indices of BRW modes is only approximately possible for much simpler structures than the ones regarded here. In our graded BRW, the spatial modes of signal and idler mostly propagate in the matching layers and are confined by the surrounding core and graded DBR layers. Following West and Helmy \cite{West--2006--Analysis-BRW}, who showed that the aluminum content of the surrounding layers affect the dispersion of the total-internal reflection and Bragg modes most strongly, we investigate the effect of these layers on the dispersion between signal and idler. As shown in \autoref{effective_index_difference}, we simulate the average DGD per length in our graded  BRW as a function of the aluminum content of the core and graded DBR layers, while keeping the other parameters unchanged. We find that a decrease of the aluminum content in the layers surrounding the matching layer  leads to a lower average DGD.

\begin{figure}
\includegraphics[width = 0.45\textwidth]{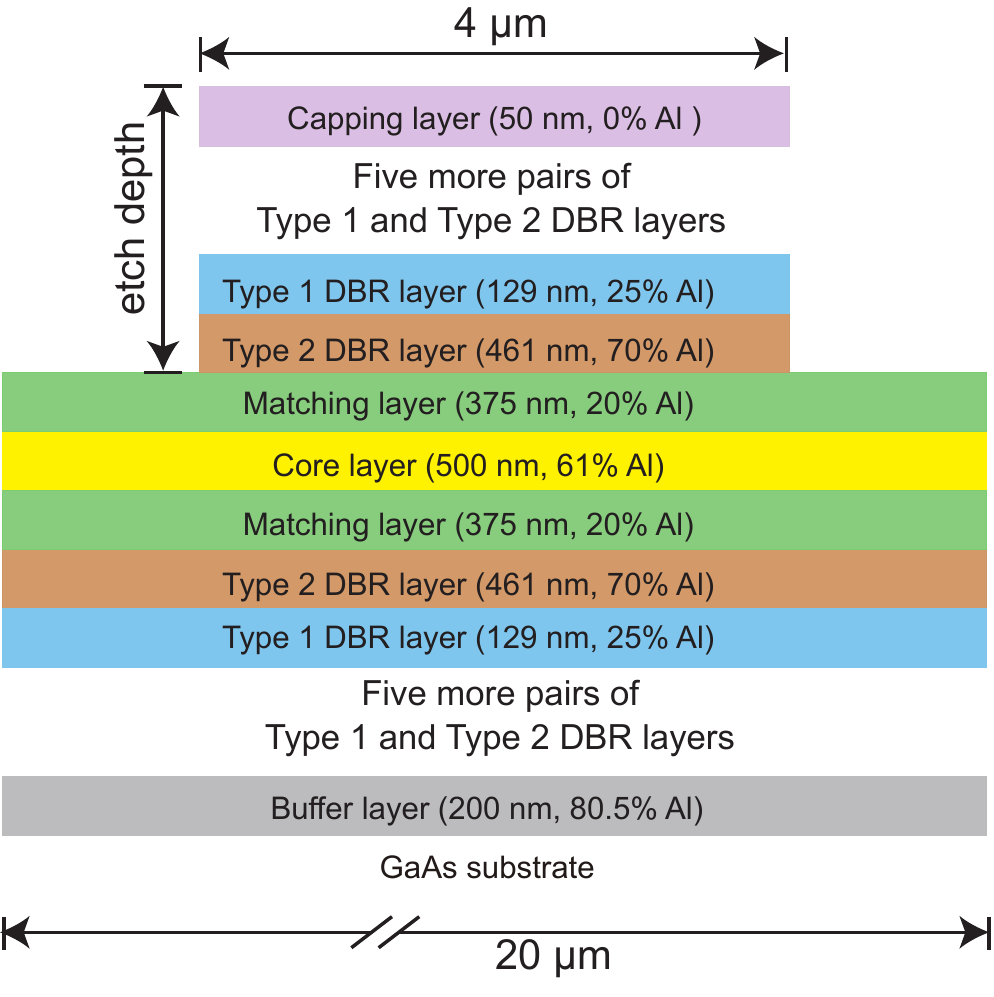}
\caption{(Color online): Layer structure of the  M-core BRW with different kinds of layers visualized with different colors. The waveguide core (yellow slab) is enclosed by the matching layers. The distributed Bragg-reflector (DBR) stacks surround this region on top and bottom. Each stack includes six layers of type 1 and type 2 DBR layers. The layer thicknesses and their aluminum contents are indicated in brackets.}
\label{fig:previous_structure}
\end{figure}

\newcolumntype{C}[1]{>{\centering\arraybackslash}p{#1}}
\begin{table}
 \begin{tabular}{C{2cm}p{0.995cm}p{0.995cm}p{0.995cm}p{0.995cm}}
  \br
Structure   & \multicolumn{1}{c}{$\tilde{n}_{\mathrm{s}} $} & \multicolumn{1}{c}{$\tilde{n}_{\mathrm{i}} $} & \multicolumn{1}{c}{$\Delta\tilde{n}$}& \multicolumn{1}{c}{$\tilde{n}_{\mathrm{p}} $} \\ \mr
 graded BRW (\autoref{fig:new_structure}) & 3.3633 & 3.3607 & 0.0026 & 4.3921 \\
 M-core BRW (\autoref{fig:previous_structure}) & 3.3385 & 3.3292 & 0.0093 & 4.0363 \\
  Ref. \cite{Gunthner-2015} & 3.33 & 3.32 & 0.01 & 4.05  \\
  Ref. \cite{laiho--2016--uncovering} & \multicolumn{2}{c}{3.31(2)} & 0.007(2) & 3.72(3) \\
  \br
\end{tabular}
\caption{Comparison of the group refractive indices for the graded  and M-core BRW structures. The group refractive indices of signal and idler and $\Delta\tilde{n} = \tilde{n}_{\mathrm{s}}-\tilde{n}_{\mathrm{i}}$ are given near \SI{1550}{\nano\meter}, whereas that of the pump mode is calculated near \SI{775}{\nano\meter}. }
\label{comparison of the dispersion}
\end{table}

Next, we compare the average DGD in the graded BRW with the one in a typical structure, the multilayer core (M-core) structure shown in \autoref{fig:previous_structure},  employed by us and others in the past \cite{Abolghasem2009_2, Gregor-Monolithic-Source-2012,Gunthner-2015,laiho--2016--uncovering}.  First, by comparing  the aluminum contents of the graded and M-core BRWs it is easy to judge  that the graded BRW has a smaller average DGD than the M-core BRW, because the aluminum concentration contrast between core and matching layers is smaller. Second, we perform a  complete simulation of the dispersion of the different modes in these two structures with commercial solvers and present in \autoref{comparison of the dispersion} the group refractive indices of the relevant modes. In the graded BRW the group index difference between signal and idler  is only $2.6\times 10^{-3}$, which corresponds to a small remaining average DGD of only \SI{4.4}{\femto\second/\milli\meter}, whereas that in the M-core BRW is 3.6 times larger.

\begin{table*}
\centering
 \begin{tabular}{p{4.0cm}p{1.1cm}p{1.1cm}p{1.1cm}p{1.1cm}p{1.1cm}p{1.1cm}}
\br
Structure&\multicolumn{2}{c}{$\kappa_{\mu}(\SI{}{\femto\second/\micro\meter})$}&\multicolumn{3}{c}{$K_{\mu}(\SI{}{\femto\second^2/\micro\meter})$} &  $\lambda_d(\SI{}{\nano\meter}) $\\ \mr
 & s & i &s& i& p &\\ \mr
graded BRW (\autoref{fig:new_structure})&$-3.429$&$-3.438$& $1.217$ & $1.181$ &$ 10.834$ &$ 1553.8$\\  
M-core BRW (\autoref{fig:previous_structure}) &$-2.326$&$-2.358$&$1.370$&$1.358  $&$ 5.068$ &$ 1550.6$\\ \br 
\end{tabular}
  \caption{Comparison of the JSA parameters for the graded  and M-core BRW designs. The parameter $\lambda_d$ denotes the designed degeneracy wavelength.}
  \label{comparison of the JSA}
\end{table*}

%----------------------------------------------------------------------------------
\section{Spectral overlap of signal and idler}
%----------------------------------------------------------------------------------

The effective group refractive indices of signal and idler not only cause a temporal delay between their wavepackets, but also considerably influence their joint spectral properties. Since the signal and idler beams from our BRWs are cross-polarized, their group refractive indices slightly differ from each other as presented in \autoref{comparison of the dispersion}. In our case this leads to spectral distinguishability between signal and idler, since their allowed frequency ranges differ from each other. Next, we study its effect.

The photon-pair state generated in the PDC process is given as \cite{Grice2001}
\begin{equation}
   |\psi\rangle = 1/\sqrt{N} \iint \mathrm{d}\omega_{s} \hspace{0.2ex}  \mathrm{d} \omega_{i} \ f(\omega_s,\omega_i) \hat{a}^{\dagger}_{\textrm{H}}(\omega_{s}) \hat{a}^{\dagger}_{\textrm{V}}(\omega_{i})\ket{0},
\label{eq:JSA}
\end{equation}
where  $f(\omega_s , \omega_i)$ is the joint spectral amplitude (JSA) expressed in terms of the signal  and idler angular frequencies, respectively, and $\hat{a}^{\dagger}_{\textrm{H,V}}$ is the photon creation operator generating photons either in horizontal (H) or vertical (V) polarization. With the help of the constant factor $N$ the JSA is normalized to $1/N\iint \mathrm{d} \omega_{s} \mathrm{d}\omega_{i} |f(\omega_s, \omega_i)| ^{2} = 1$. Due to energy conservation,  $\omega_s +\omega_i = \omega_p$  is maintained in the PDC process with $ \omega_p$ denoting the angular frequency of pump.

Following Refs~\cite{Gunthner-2015,laiho--2016--uncovering} we write the JSA as a product of the pump beam spectral envelope $\alpha(\omega_p) = \exp(- (\omega_{p} -\varpi)^{2}/\sigma^{2})$, in which  $\sigma$  is related to  the pump bandwidth and $\varpi$  denotes its central angular frequency, and the phasematching function $ \phi(\omega_{s},\omega_{i}) = \mathrm{sinc} \left( \Delta k (\omega_s, \omega_i)\ L/2 \right)e^{-i \Delta k(\omega_s, \omega_i)\ L/2} $, in which  $\Delta k (\omega_s,\omega_i) = k_{s}(\omega_{s}) +k_{i}(\omega_{i}) -k_{p}(\omega_{p}) $ accounts for the phase mismatch. We describe it in terms of the propagation constants $k_{\mu}(\omega_{\mu}) = n_{\mu}(\omega_{\mu})  \omega_{\mu}/c$ ($\mu= p,s,i$) of pump, signal, and idler.  Moreover, we investigate the JSA around a phasematched frequency triplet $\omega^{0}_{s} +\omega^{0}_{i} = \omega^{0}_{p} $, for which $\Delta k(\omega^{0}_{s}, \omega^{0}_{i})  = 0$. Thus, we re-write the phase mismatch in  terms of the detunings $\nu_{\mu} = \omega_{\mu} - \omega^{0}_{\mu}$, and  approximate it as 
   \begin{align}
    \Delta k \approx & \phantom{0} \kappa_{s}\nu_s +\kappa_{i} \nu_i \nonumber \\
 &+1/2(K_s-K_p)\nu^{2}_{s}+1/2(K_i-K_p)\nu^{2}_{i} -K_p\nu_{s}\nu_{i},   \nonumber
\end{align}
in which  $\kappa_{\mu}$$= 1/c\left[\tilde{n}_{\mu}(\omega^{0}_{\mu})- \tilde{n}_{p}(\omega^{0}_{p}) \right]$ and $K_{\mu} = 1/c  \frac{\textrm{d} \tilde{n}_{\mu}(\omega)}{\textrm{d}\omega} |_{\omega = \omega^{0}_{\mu}}$ are related to the effective group refractive indices of the interacting modes.

\begin{figure*}
  \centering
  \includegraphics[width=0.96\textwidth]{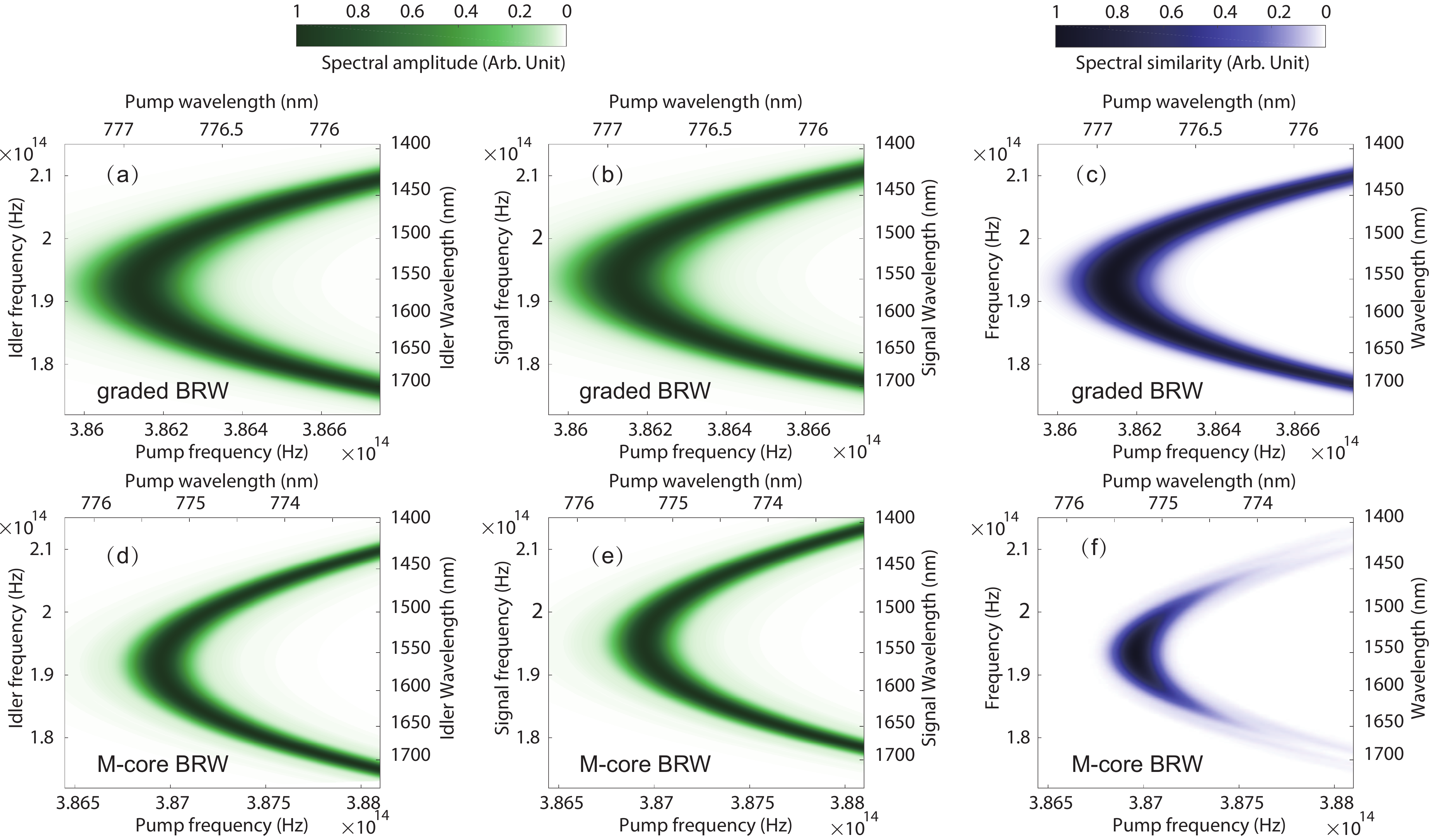}
  \caption{Calculated single photon spectrum of (a) signal and (b) idler for the graded BRW structure as well as (c) their product for visualizing the similarity of the generated spectral bands. In (d-f) we present the same for the M-core BRW structure. }
  \label{fig:spectrum}
\end{figure*}

In \autoref{comparison of the JSA} we list the phasematching parameters for the two investigated BRW designs. With these parameters we simulate the photon-pair characteristics assuming a Gaussian pump amplitude with a full width at half maximum (FWHM) of \SI{0.25}{\nano\meter} and a waveguide length of \SI{2}{\milli\meter}. We keep these parameters the same in all the following simulations. For the graded  BRW structure we illustrate in \autoref{fig:spectrum} (a-b) the single photon spectra of signal and idler, respectively, as well as in \autoref{fig:spectrum} (c) their product for a better visualization of the similarity of their spectral extent. In \autoref{fig:spectrum} (d-f) we present the same for the M-core BRW structure. Since the  spectral extent of signal and idler depends strongly on the group refractive indices of the interacting modes, a comparison between \autoref{fig:spectrum}(c) and (f) clearly  reveals the differences in the optical properties of PDC photons emitted from the two investigated BRW structures.

\begin{figure}[H]
\begin{minipage}[t]{\linewidth}
   \includegraphics[width=0.95\textwidth]{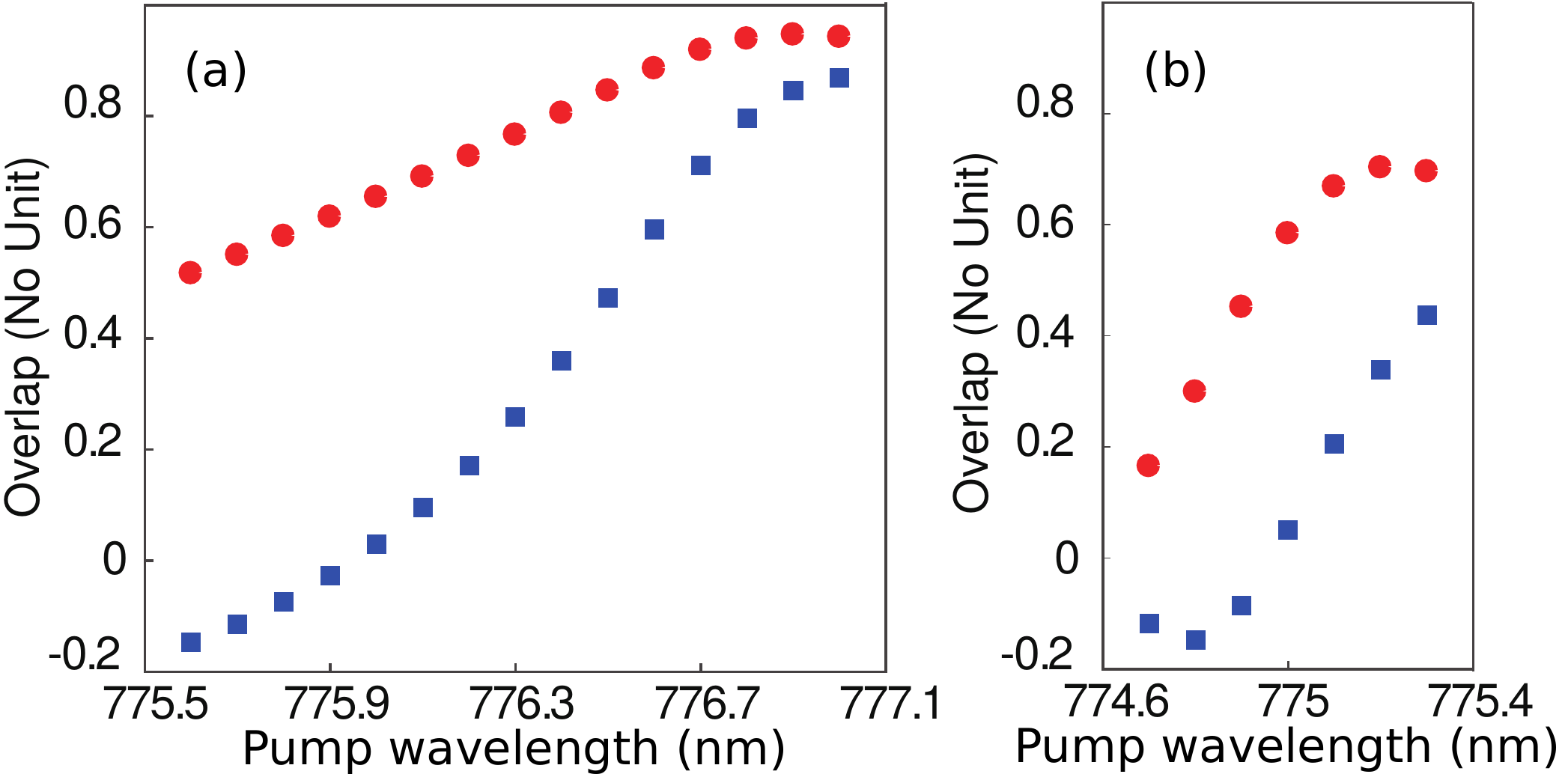}
\end{minipage}
\caption{The overlap in the (a) graded  and (b) M-core BRW structures calculated via \autoref{eq:overlap} in terms of the pump wavelength. The red circles illustrate the overlap between signal and idler at $\tau = 0$, while the blue squares represent the overlap achieved, when the signal and idler wavepackets are delayed with $\tau = \tau_{c}$ that is \SI{-9}{\femto\second} in (a) and \SI{-31}{\femto\second} in (b). }
\label{fig:overlap_tau}
\end{figure}

In order to investigate the spectro-temporal indistinguishability of signal and idler, we calculate their
overlap given by \cite{W.P.Grice1997, Avenhaus2009}
\begin{equation}
 \mathcal{O}(\tau) = \frac{\iint  \mathrm{d} \omega_s  \mathrm{d} \omega_i f(\omega_s,\omega_i)f^*(\omega_i,\omega_s)e^{i(\omega_s-\omega_i)\tau}}{\iint \mathrm{d} \omega_s \mathrm{d} \omega_i |f(\omega_s,\omega_i)|^2}
\label{eq:overlap}
\end{equation}
in terms of a temporal delay  $\tau$, which can be introduced by retarding the signal and idler wavepackets relative to each other. The overlap measures the symmetry of the JSA with respect to the exchange of signal and idler frequencies and includes also a temporal compensation. In \autoref{fig:overlap_tau}(a)  we present the overlap values for the graded BRW structure and in   \autoref{fig:overlap_tau}(b)  for the M-core BRW structure with respect to the central wavelength of the pump beam. The overlap reaches a value as high as 85\% (44\%) for the graded  (M-core) BRW structure without any temporal compensation. If the optimal temporal compensation $|\tau_{c}| = \textrm{DGD}$ of approximately \SI{9}{\femto\second} (\SI{31}{\femto\second}) is applied, the value of the overlap increases to its maximal value $\mathcal{O}_{\textrm{max.}}$ of 94.6\% (69.5\%). When detuning the pump towards shorter wavelengths, the overlap gradually decreases for both investigated BRWs. However, the decay is faster in the M-core BRW structure due to the worse spectral similarity. The overlap value can even become negative if the temporal delay between signal and idler is not compensated. This effect can be taken  advantage of when investigating the  bunching of photon pairs next in Section  \uppercase\expandafter{5}.

%----------------------------------------------------------------------------------
\section{BRWs in quantum optics applications}
%----------------------------------------------------------------------------------

The spectro-temporal properties of photon pairs play a crucial role when adapting a PDC source to a quantum optics task. Therefore, we compare the performance of both investigated BRW structures in such applications. We start by exploring their suitability for a HOM quantum interference experiment \cite{Zhukovsky2013, Jin2018, Graffitti2018} and thereafter study the preparation of spectrally multi-band polarization entanglement     \cite{kang2016monolithic,Claire2016Multi-user,Chen2017}.

\begin{figure}
  \centering
  \includegraphics[width=0.4\textwidth]{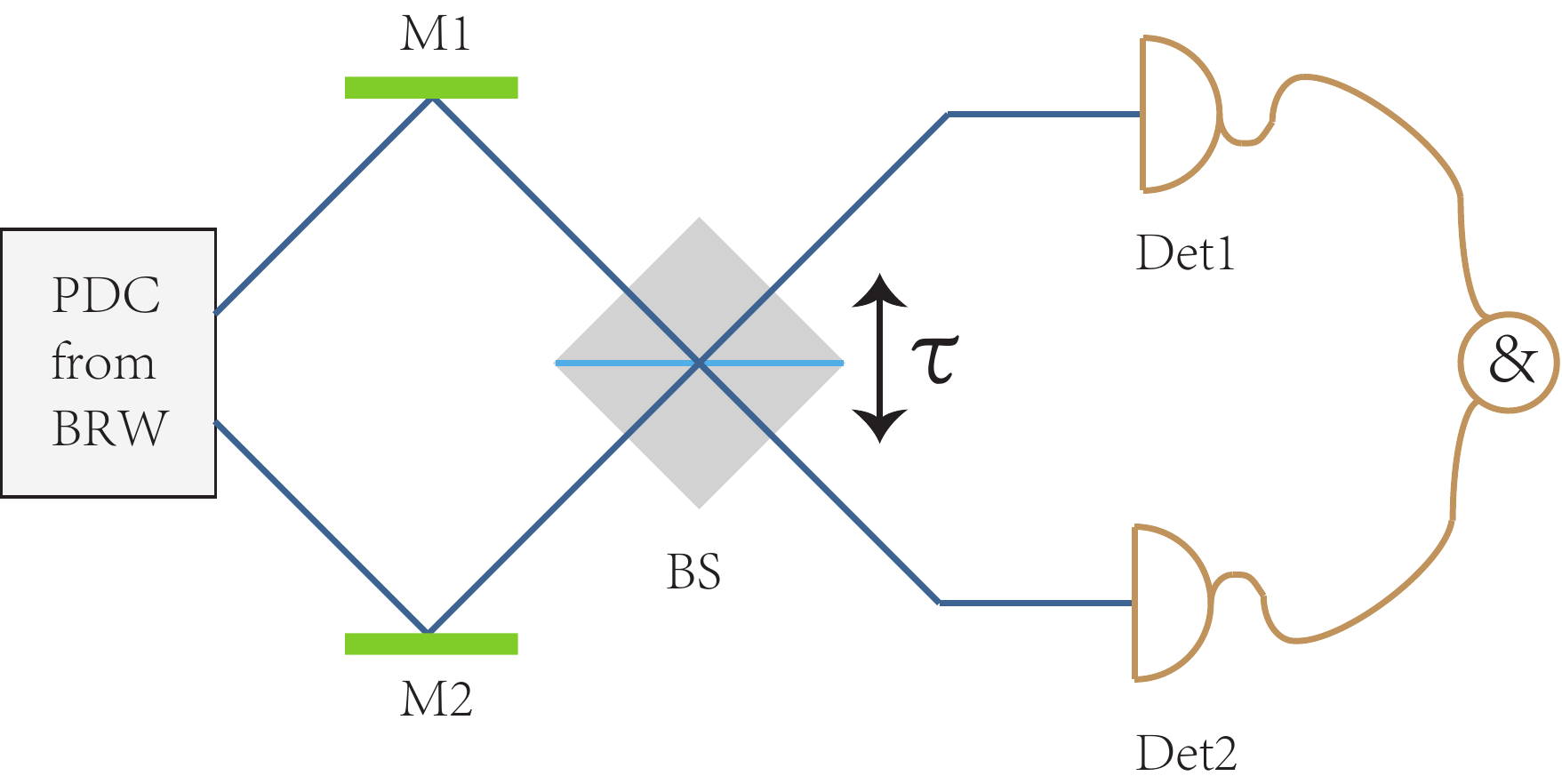}
  \caption{Schematic of a HOM experiment. The parameter $\tau$ is the delay between the two arms. Abbreviations: M: mirror; BS: beam splitter; Det: single-photon sensitive on/off-detector. The symbol \& denotes coincidence discrimination.}
  \label{HOM_scheme}
\end{figure}

\begin{figure}
\centering
\includegraphics[width=0.48\textwidth]{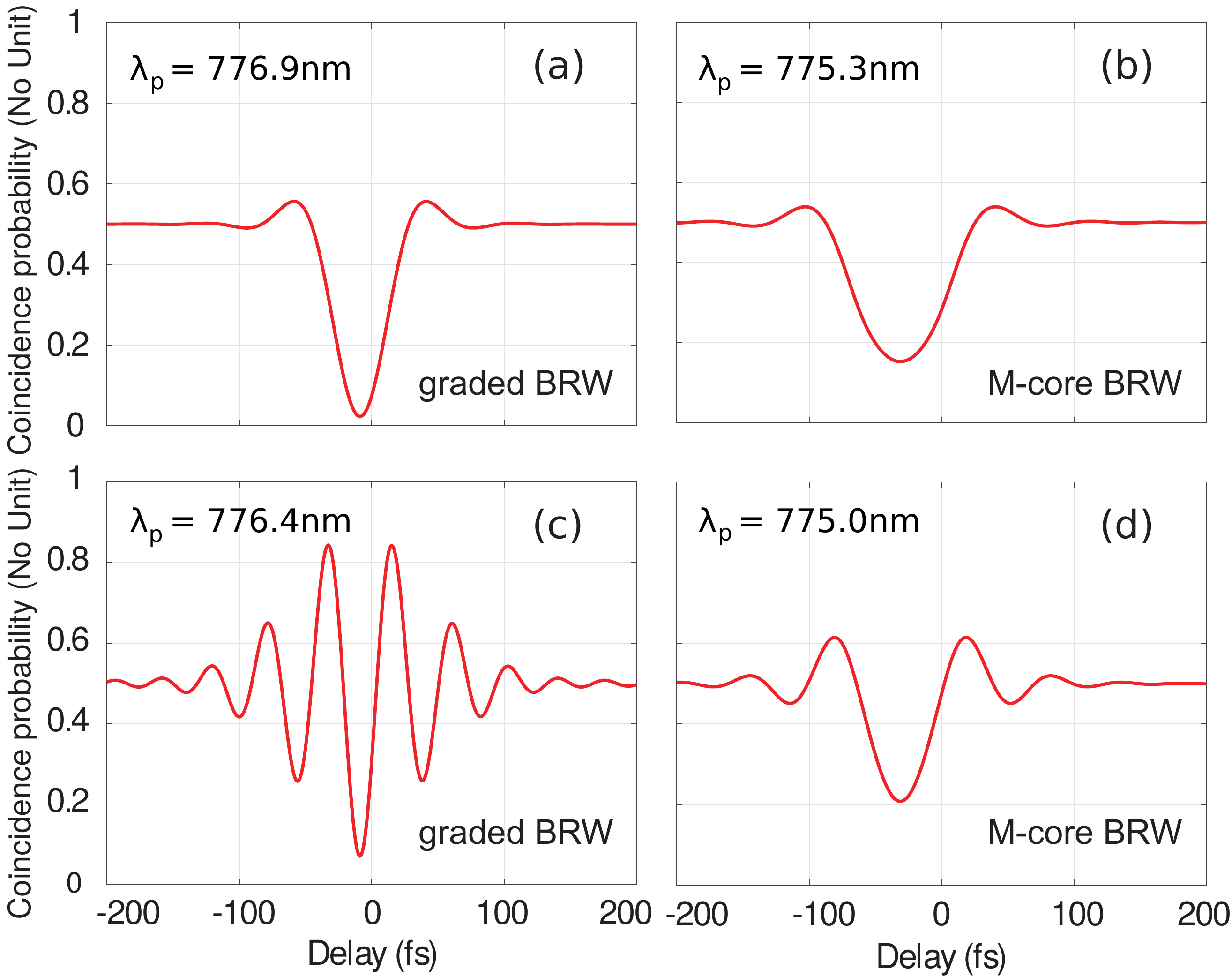}
\caption{HOM interference in the two investigated BRW structures. The comparison of the HOM dips at the degeneracy for (a) the  graded   and (b) M-core BRW structures without any filters reveals the clear difference in the spectral indistinguishability of signal and idler. In (c) and (d) we illustrate the coincidence click probabilities without any filters for the graded  and M-core BRW structures when the pump wavelength is moved to   \SI{776.4}{\nano\meter}  and \SI{775.0}{\nano\meter}, respectively.}
\label{HOM_fringe}
\end{figure}

In the HOM interference experiment \cite{C.K.Hong1987} as sketched in \autoref{HOM_scheme} the probability $P$ of measuring a coincidence click  between the output ports of a symmetric beam splitter, when sending a photon pair from PDC to its input ports, is directly connected to the overlap in \autoref{eq:overlap} and is given by \cite{W.P.Grice1997}
\begin{equation}
P = \frac{1}{2}-\frac{1}{2}\mathcal{O}(\tau)
\label{eq:Phom}
\end{equation}
assuming that our BRW emits PDC with only a small amount of spurious noise such that the background from higher-photon number contributions  can be neglected \cite{Chen2018}.
We simulate the HOM interference for the two investigated BRW structures at their respective degeneracy wavelengths and far away from them. In \autoref{HOM_fringe}(a) and (b) we illustrate the HOM dips at the degeneracy, that is, the PDC processes in the graded and M-core BRW structures are pumped at the wavelengths of \SI{776.9}{\nano\meter} and \SI{775.3}{\nano\meter}, respectively. Clearly, the graded BRW  structure can produce photon pairs with a higher indistinguishability than the M-core BRW. By utilizing \autoref{eq:Phom} the HOM dip visibility given by $[P(\tau \rightarrow\infty)-P(\tau = \tau_{c})]/P(\tau \rightarrow\infty)$ reduces to $\mathcal{O}_{\textrm{max.}}$ for the cases in \autoref{HOM_fringe}(a) and (b) and takes the values reported in Section \uppercase\expandafter{4}. When detuning the central wavelength of the pump towards shorter wavelengths, the HOM interference starts showing fringes, if signal and idler are generated in two separate spectral regions \cite{Eckstein2008}. In  \autoref{HOM_fringe}(c) we show this HOM interference pattern for the graded BRW structure at a pump wavelength of \SI{776.4}{\nano\meter}, which illustrates that we can utilize the material dispersion to control the HOM dip characteristics without spectral filtering. However, this is not the case with the M-core BRW structure, in which the visibility of the HOM dip drops as we tune the pump wavelength to \SI{775.0}{\nano\meter} and hardly shows any fringes as illustrated in \autoref{HOM_fringe}(d). 

\begin{figure}
  \centering
  \includegraphics[width=0.4\textwidth]{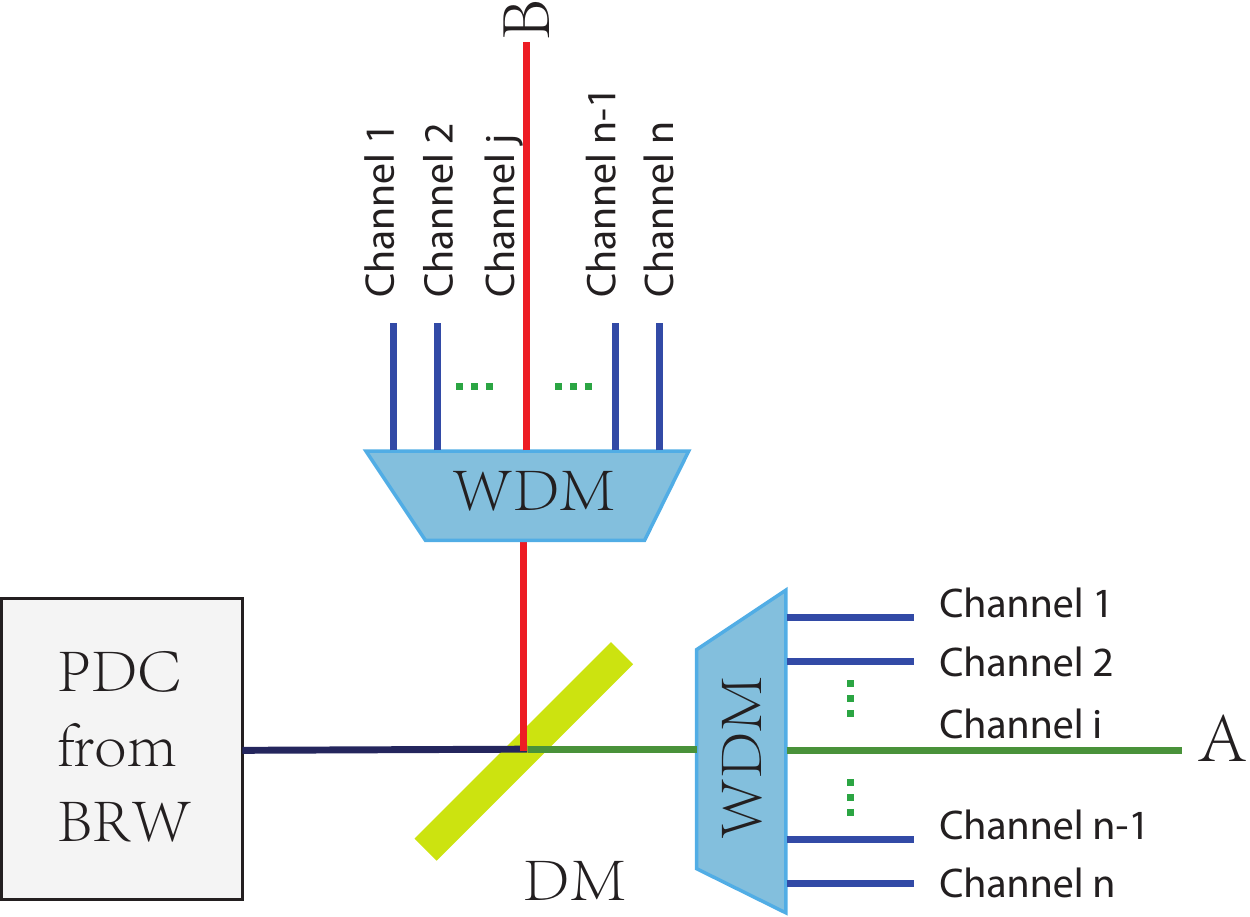}
  \caption{Scheme for generating polarization entangled states between the paths A and B. Abbreviations: DM: dichroic mirror; WDM: wavelength division multiplexer.}
  \label{polarization_entangled_states_generation_scheme}
\end{figure}

Finally, we use the scheme in \autoref{polarization_entangled_states_generation_scheme} for generating polarization entangled states with multiple spectral bands that are \textit{free from temporal delay compensation} \cite{kang2016monolithic,Claire2016Multi-user,Chen2017}. In the absence of background light, the density matrix of the polarization entangled state takes the form \cite{schlager--2017--temporally},
\begin{equation}
\begin{split}
  \rho & = \alpha\ket{\textrm{HV}}\bra{\textrm{HV}} + \mathcal{D}\ket{\textrm{VH}}\bra{\textrm{HV}}\\
       & + \mathcal{D}^{\ast}\ket{\textrm{HV}}\bra{\textrm{VH}} + \beta\ket{\textrm{VH}}\bra{\textrm{VH}},\\
\end{split}
\label{eq:density_matrix}
\end{equation}
in which $\ket{\textrm{H}}$ and $\ket{\textrm{V}}$ describe a single-photon state with horizontal polarization (signal) and vertical polarization (idler), respectively.  In \autoref{eq:density_matrix} the diagonal elements take the form $\alpha = 1/\mathcal{N}\iint \mathrm{d}\omega_s\mathrm{d}\omega_i|g(\omega_s,\omega_i)|^2$ and $\beta  = 1/\mathcal{N}\iint \mathrm{d}\omega_s\mathrm{d}\omega_i|h(\omega_s,\omega_i)|^2$ with  $\mathcal{N}$ being a normalization constant, whereas the off-diagonal elements are described as
\begin{equation}
 \mathcal{D}  = 1/\mathcal{N}\iint \mathrm{d}\omega_s\mathrm{d}\omega_{i}h(\omega_i,\omega_s)g^{\ast}(\omega_s,\omega_i)
\end{equation}
with
\begin{equation}
\begin{split}
  g(\omega_s,\omega_i)  = f(\omega_s,\omega_i)G_1(\omega_s)G_2(\omega_i)\sqrt{T(\omega_s)R(\omega_i)},\\
\end{split}
\label{eq:g_function}
\end{equation}
and
\begin{equation}
\begin{split}
  h(\omega_s,\omega_i)  = f(\omega_s,\omega_i)G_1(\omega_i)G_2(\omega_s)\sqrt{T(\omega_i)R(\omega_s)}.\\
\end{split}
\label{eq:h_function}
\end{equation}
 In \autoref{eq:g_function} and \autoref{eq:h_function} $G_1(\omega) = \textrm{e}^{-\frac{(\omega - \omega_1)^2}{\sigma^{2}_1}}$ and $G_2(\omega) = \textrm{e}^{-\frac{(\omega - \omega_2)^2}{\sigma^{2}_2}}$ describe the amplitudes of Gaussian bandpass filters with central angular frequencies  $\omega_1$ and $\omega_2$ and bandwidths $\sigma_1$ and $\sigma_2$, respectively, while $T(\omega)$ and $R(\omega)$ are the transmittance and reflectance  of the dichroic mirror in \autoref{polarization_entangled_states_generation_scheme}, for which $T(\omega)+ R(\omega) = 1$. For simplicity, we take $T(\omega)$ and $R(\omega)$  as step functions having the cut-off wavelength at the degeneracy.

\begin{figure}
\centering
\includegraphics[width=0.45\textwidth]{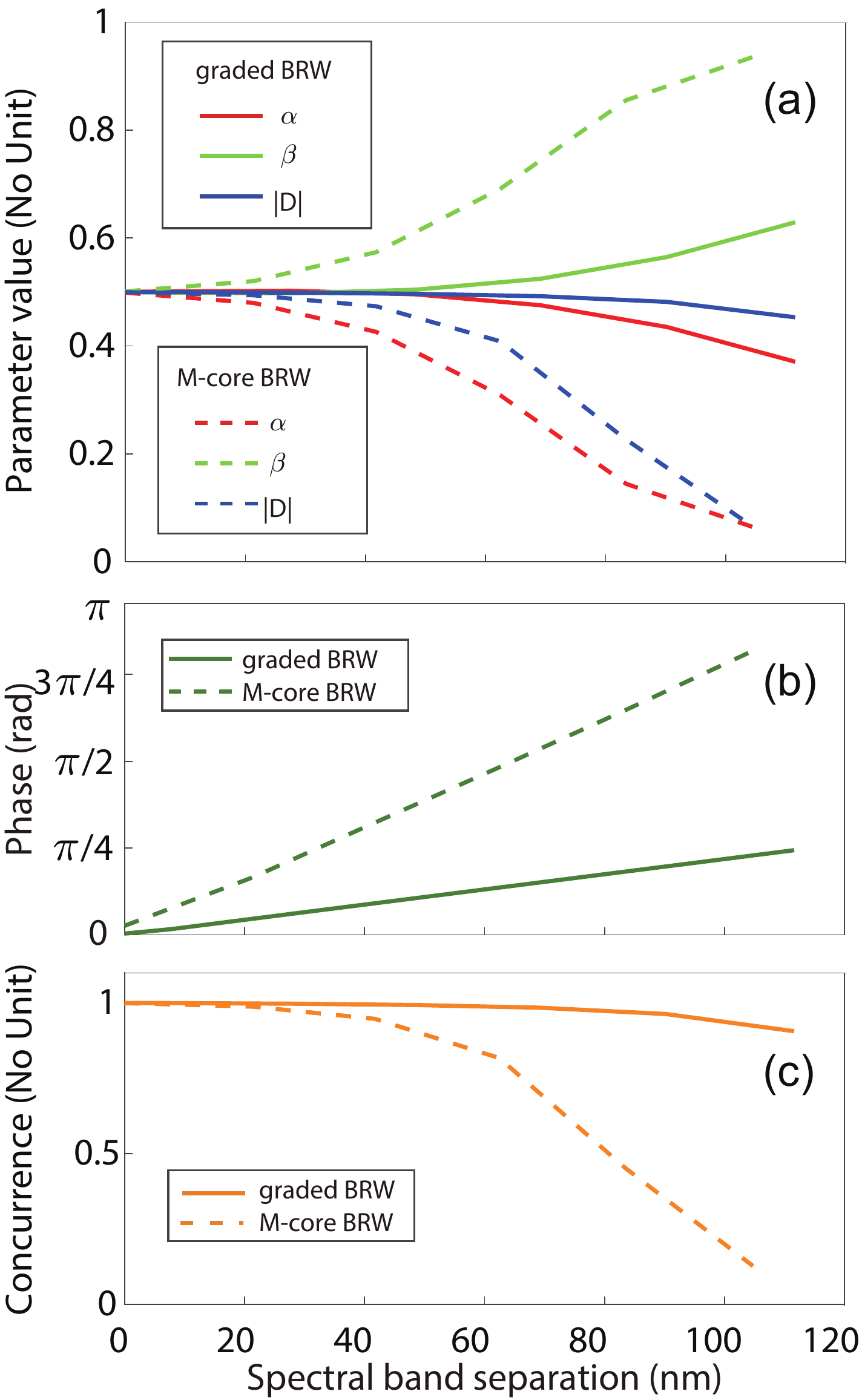}
\caption{Simulation of the properties of the density matrix in \autoref{eq:density_matrix} for the two investigated  BRW structures.
(a) The simulated values of $\alpha$, $\beta$ and $|\mathcal{D}|$, (b) the phase of the D-parameter ($|\textrm{arg}(\mathcal{D})|$) and (c) the concurrence are illustrated with respect to the spectral band separation in the paths A and B in \autoref{polarization_entangled_states_generation_scheme}. }
  \label{D_parameter}
\end{figure}

Due to the normalization, the diagonal elements of the density matrix in \autoref{eq:density_matrix} obey the relation  $\alpha +\beta = 1$, while the off-diagonal elements $\mathcal{D}$ and $\mathcal{D}^{\ast}$, called coherences, can be used to quantify the amount of entanglement. Ideally, for a maximally entangled state $\alpha = \beta = |\mathcal{D}| = 1/2$. Thus, the behavior of the $\mathcal{D}$-parameter  is especially interesting, because any decrease in its value means a reduced  amount of created entanglement.

\begin{figure}
\begin{minipage}[t]{1\linewidth}
  \centering
  \includegraphics[width=0.99\textwidth]{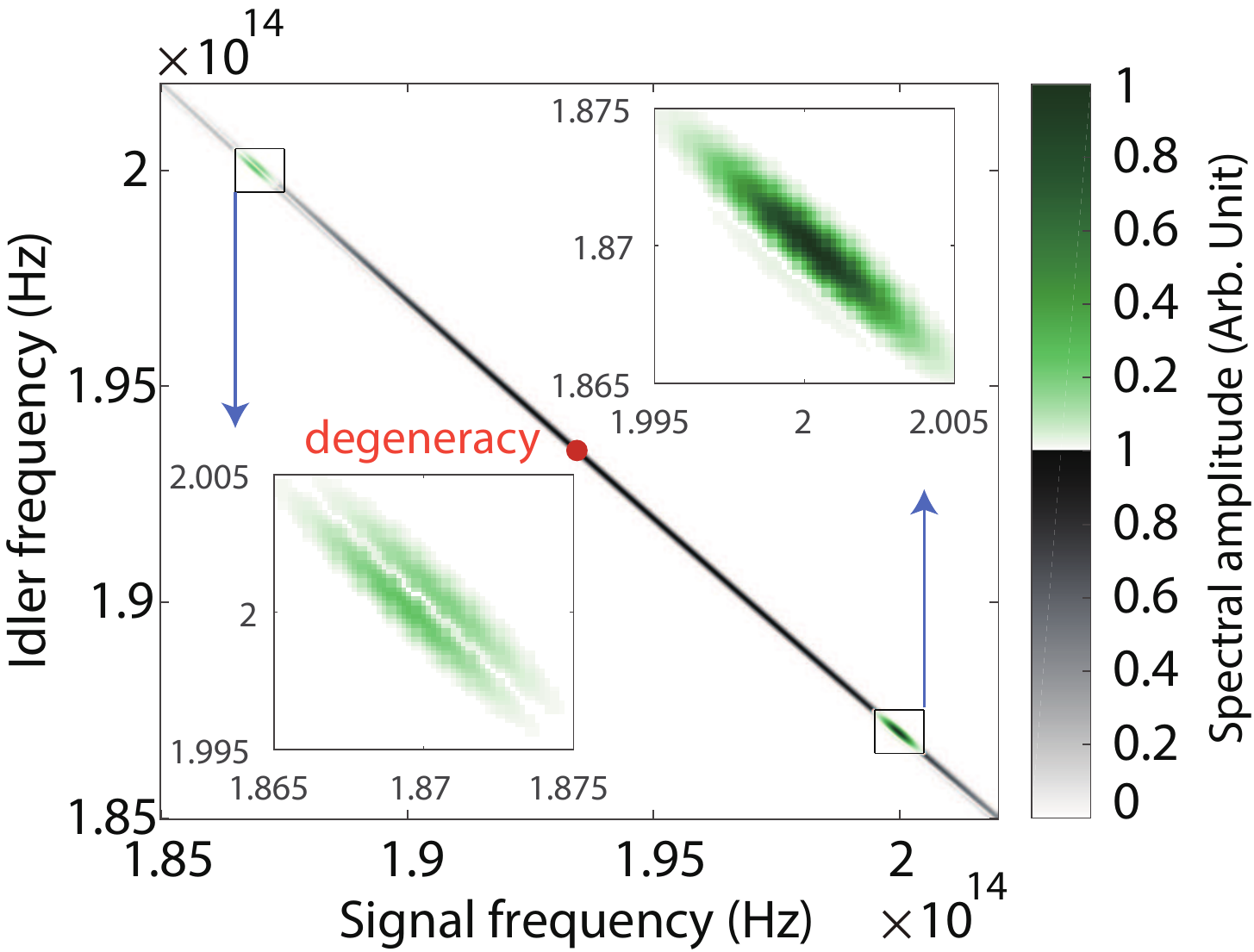}
\end{minipage}
  \caption{(Color online:) The unfiltered JSA of the M-core BRW structure when pumped at \SI{775.3}{\nano\meter} (gray contour) together with the JSA after the PDC emission is filtered around \SI{1500}{\nano\meter} and \SI{1604.7}{\nano\meter} (green countours).  The insets show a zoom in the filtered JSA. The  upper and lower frequency bands of the filtered JSA clearly have different weights causing imperfections in the density matrix elements of the polarization entangled state. The red dot marks the degeneracy.}
  \label{JSA_old}
\end{figure}

In \autoref{D_parameter} we show the density matrix elements from  \autoref{eq:density_matrix}  for the two investigated BRW structures with respect to the separation $|\omega_{2}-\omega_{1}|$ of the signal and idler spectral bands.  We simulate the PDC processes at the degeneracy and and use narrowband spectral Gaussian filters with a FWHM of \SI{2}{\nano\meter} for creating entangled photon pairs in different spectral bands obeying the energy conservation $\omega_{1}+\omega_{2} = \omega_{p}$.
 From \autoref{D_parameter}(a)  we see that the density matrix elements calculated for the graded  BRW structure are more robust against a large spectral band separation  than those calculated for the M-core BRW structure. Additionally, the phase of the $\mathcal{D}$-parameter in \autoref{D_parameter}(b) is varying less for the graded  BRW structure than for the  M-core BRW structure. This indicates that the entangled states generated in different frequency bands have more uniform characteristics if the average DGD is small. Finally, in \autoref{D_parameter}(c)  we quantify the achievable entanglement in both BRW structures  with the concurrence \cite{Hill1997}.

Apart from the temporal walk-off of signal and idler, also their spectral overlap plays a crucial role in the generation of polarization entanglement. The asymmetry of the JSA with respect to the degeneracy  as shown in \autoref{JSA_old} for the M-core BRW structure is the reason for the highly imbalanced diagonal elements $\alpha$ and $\beta$ in \autoref{D_parameter}(a) and the low concurrence in  \autoref{D_parameter}(c). To summarize, our results show that the layer parameters of BRWs  have to be carefully designed in order to optimize their performance in  quantum optical applications.

%----------------------------------------------------------------------------------
\section{Conclusion}
%----------------------------------------------------------------------------------

Direct engineering of the spectro-temporal properties of PDC emission is of great importance for preparing quantum optical states with high quality. We investigated how the shape and the material composition of a Bragg-reflection waveguide affect  its optical properties, like the phasematching wavelength and the group indices of the interacting modes and showed that our design is to a well-manageable degree tolerant against variations. We further investigated, how the effective group refractive index difference between signal and idler can be manipulated. Our BRW design results in good spectral and temporal overlap between signal and idler, which makes our structure well-suited for different quantum optics tasks both at the degeneracy and far away from it.

%----------------------------------------------------------------------------------
%                Acknowledgements
%----------------------------------------------------------------------------------
\section*{Acknowledgements}
This work was supported by the Austrian Science Fund (FWF) through the project nos.~I-2065 and J-4125, the DFG project no. SCHN1376/2-1, the State of Bavaria, China Scholarship Council project no. 201503170272, and Natural Science Foundation of Hunan Province of China project no. 2018JJ2467.

\section*{References}
\providecommand{\noopsort}[1]{}\providecommand{\singleletter}[1]{#1}%


\begin{thebibliography}{10}

\bibitem{caspani2017integrated}
L. Caspani, C. Xiong, B.~J. Eggleton, D. Bajoni, M. Liscidini, M. Galli, R. Morandotti, and D.~J. Moss.
\newblock Integrated sources of photon quantum states based on nonlinear optics.
\newblock {\em Light Sci. Appl.}, 6:e17100, 2017.

\bibitem{Flamini2018}
F.~Flamini, N.~Spagnolo, and F. Sciarrino.
\newblock Photonic quantum information processing: a review.
\newblock {\em  Rep. Prog. Phys},. 82 016001, 2019.

\bibitem{C.K.Hong1987}
C.~K. Hong, Z.~Y. Ou, and L.~Mandel.
\newblock Measurement of subpicosecond time intervals between two photons by interference.
\newblock {\em Phys.~Rev.~Lett.}, 59:2044, 1987.

\bibitem{Evans2010}
P.~G. Evans, R.~S. Bennink, W.~P. Grice, T.~S. Humble, and J.~Schaake.
\newblock Bright source of spectrally uncorrelated polarization-entangled photons with nearly single-mode emission.
\newblock {\em Phys. Rev. Lett.}, 105:253601, 2010.

\bibitem{Zhu2012}
E.~Y. Zhu, Z.~Tang, L.~Qian, L.~G. Helt, M.~Liscidini, J.~E. Sipe, C.~Corbari, A.~Canagasabey, M.~Ibsen, and P.~G. Kazansky.
\newblock Direct generation of polarization-entangled photon pairs in a poled fiber.
\newblock {\em Phys. Rev. Lett.}, 108:213902, 2012.

\bibitem{Bruno2014a}
N.~Bruno, A.~Martin, T.~Guerreiro, B.~Sanguinetti, and R.~T. Thew.
\newblock Pulsed source of spectrally uncorrelated and indistinguishable  photons at telecom wavelengths.
\newblock {\em Opt. Express}, 22:17246, 2014.

\bibitem{Kang2014}
D.~Kang, A.~Pang, Y., Zhao, and A.~S. Helmy.
\newblock Two-photon quantum state engineering in nonlinear photonic nanowires.
\newblock {\em J. Opt. Soc. Am. B}, 31:1581, 2014.

\bibitem{Laiho2011a}
K.~Laiho, A.~Christ, K.~N. Cassemiro, and Ch. Silberhorn.
\newblock Testing spectral filters as gaussian quantum optical channels.
\newblock {\em Opt. Lett.}, 36:1476, 2011.

\bibitem{Bruno2014}
N.~Bruno, E.~Z. Cruzeiro, A.~Martin, and R.~T. Thew.
\newblock Simple, pulsed, polarization entangled photon pair source.
\newblock {\em Opt. Comm.}, 327:3, 2014.

\bibitem{Silverstone2014}
J.~W. Silverstone, D.~Bonneau, K.~Ohira, N.~Suzuki, H.~Yoshida, N.~Iizuka,  M.~Ezaki, C.~M. Natarajan, M.~G. Tanner, R.~H. Hadfield, V.~Zwiller, G.~D.  Marshall, J.~G. Rarity, J.~L. O'Brien, and M.~G. Thompson.
\newblock On-chip quantum interference between silicon photon-pair sources.
\newblock {\em Nature Photon.}, 8:104, 2014.

\bibitem{Donohue2016}
J.~M. Donohue, M.~Mastrovich, and K.~J. Resch.
\newblock Spectrally engineering photonic entanglement with a time lens.
\newblock {\em Phys. Rev. Lett.}, 117:243602, 2016.

\bibitem{Wright2017}
L.~J. Wright, M.~Karpinski, C.~S\"{o}ller, and B.~J. Smith.
\newblock Spectral shearing of quantum light pulses by electro-optic phase modulation.
\newblock {\em Phys. Rev. Lett.}, 118:023601, 2017.

\bibitem{Puigibert2017}
M. Grimau Puigibert, G. H. Aguilar, Q. Zhou, F. Marsili, M. D. Shaw, V. B. Verma, S. W. Nam, D. Oblak, and W. Tittel.
\newblock Heralded single photons based on spectral multiplexing and  feed-forward control.
\newblock {\em Phys. Rev. Lett.}, 119:083601, 2017.

\bibitem{Ansari2018}
V.~Ansari, E.~Roccia, M.~Santandrea, M.~Doostdar, C.~Eigner, L.~Padberg,  I.~Gianani, M.~Sbroscia, J.~M. Donohue, L.~Mancino, M.~Barbieri, and C.~Silberhorn.
\newblock Heralded generation of high-purity ultrashort single photons in programmable temporal shapes.
\newblock {\em Opt. Express}, 26:2764, 2018.

\bibitem{Zhukovsky2012}
S.~V. Zhukovsky, L.~G. Helt, D.~Kang, P.~Abolghasem, A.~S. Helmy, and J.~E. Sipe.
\newblock Generation of maximally-polarization-entangled photons on a chip.
\newblock {\em Phys. Rev. A}, 85:013838, 2012.

\bibitem{Horn-scientific-reports-2013}
R.~T Horn, P. Kolenderski, D. Kang, P. Abolghasem, C. Scarcella, A. Della~Frera, A. Tosi, L.~G. Helt, S.~V.  Zhukovsky, J.~E. Sipe, et~al.
\newblock Inherent polarization entanglement generated from a monolithic semiconductor chip.
\newblock {\em Sci. Rep.}, 3, 2013.

\bibitem{kang2016monolithic}
D. Kang, A. Anirban, and A.~S. Helmy.
\newblock Monolithic semiconductor chips as a source for broadband wavelength-multiplexed polarization entangled photons.
\newblock {\em Opt. Express}, 24:15160, 2016.

\bibitem{Claire2016Multi-user}
C. Autebert, J. Trapateau, A. Orieux, A. Lema{\^\i}tre,  C. Gomez-Carbonell, E. Diamanti, I. Zaquine, and S. Ducci.
\newblock Multi-user quantum key distribution with entangled photons from an AlGaAs chip.
\newblock {\em Quantum Sci. Tech.}, 1:01LT02, 2016.

\bibitem{schlager--2017--temporally}
A.~Schlager, B.~Pressl, K.~Laiho, H.~Suchomel, M.~Kamp, S.~H\"{o}fling, C.~Schneider,  and G.~Weihs.
\newblock Temporally versatile polarization entanglement from Bragg reflection waveguides.
\newblock {\em Opt. Lett.}, 42:2102, 2017.

\bibitem{lanco2006semiconductor}
L.~Lanco, S.~Ducci, J.-P. Likforman, X.~Marcadet, J. A. W. Van~Houwelingen, H. Zbinden,  G.~Leo, and V.~Berger.
\newblock Semiconductor waveguide source of counterpropagating twin photons.
\newblock {\em Phys. Rev. Lett.}, 97:173901, 2006.

\bibitem{sarrafi2013continuous}
P. Sarrafi, E.~Y. Zhu, K. Dolgaleva, B.~M. Holmes, D.~C.  Hutchings, J.~S. Aitchison, and L.~Qian.
\newblock Continuous-wave quasi-phase-matched waveguide correlated photon pair source on a III--V chip.
\newblock {\em Appl. Phys. Lett.}, 103:251115, 2013.

\bibitem{Gregor-Monolithic-Source-2012}
R. Horn, P. Abolghasem, B.~J. Bijlani, D. Kang, A.~S. Helmy, and G. Weihs.
\newblock Monolithic source of photon pairs.
\newblock {\em Phys. Rev. Lett.}, 108:153605, 2012.

\bibitem{West--2006--Analysis-BRW}
B.~R. West and A.~S. Helmy.
\newblock Analysis and design equations for phase matching using Bragg reflector waveguides.
\newblock {\em IEEE J. Quantum Electron.}, 12:431, 2006.

\bibitem{Helmy--2006--Phase-matching}
A.~S. Helmy.
\newblock Phase matching using Bragg reflection waveguides for monolithic  nonlinear optics applications.
\newblock {\em Opt. Express}, 14:1243, 2006.

\bibitem{Gehrsitz-refractive-index-2000}
S.~Gehrsitz, F. K.~Reinhart, C.~Gourgon, N.~Herres, A.~Vonlanthen, and H.~Sigg.
\newblock The refractive index of Al$_{x}$Ga$_{1- x}$As below the band gap: accurate determination and empirical modeling.
\newblock {\em J. Appl. Phys.}, 87:7825, 2000.

\bibitem{Pressl--2015}
B.~Pressl, T.~G\"{u}nthner, K.~Laiho, J.~Ge{\ss}ler, M.~Kamp, S.~H\"{o}fling,  C.~Schneider, and G.~Weihs.
\newblock Mode-resolved fabry-perot experiment in low-loss Bragg-reflection  waveguides.
\newblock {\em Opt. Express}, 23:33608, 2015.

\bibitem{laiho--2016--uncovering}
K.~Laiho, B.~Pressl, A.~Schlager, H.~Suchomel, M.~Kamp, S.~H{\"o}fling, C.~Schneider, and G.~Weihs.
\newblock Uncovering dispersion properties in semiconductor waveguides to study photon-pair generation.
\newblock {\em Nanotechnology}, 27:434003, 2016.

\bibitem{Misiaszek--2018}
M. Misiaszek, A. Gajewski and P. Kolenderski
\newblock Dispersion measurement method with down conversion process.
\newblock {\em J. Phys. Commun.} 2:065014, 2018.

\bibitem{Pressl-2018-Semi-automatic}
B.~Pressl, K.~Laiho, H.~Chen, T.~G\"unthner, A.~Schlager, S.~Auchter, H.~Suchomel, M.~Kamp, S.~H\"{o}fling, C.~Schneider, and G.~Weihs.
\newblock Semi-automatic engineering and tailoring of high-efficiency Bragg-reflection waveguide samples for quantum photonic applications.
\newblock {\em Quantum Sci. Tech.}, 3:024002, 2018.

\bibitem{Comsol}
COMSOL Multiphysics\circledR, \url{www.comsol.com}, COMSOL AB, Stockholm, Sweden.

\bibitem{Bijlani2008}
B. Bijlani, P. Abolghasem,  and A. S. Helmy
\newblock Second harmonic generation in ridge Bragg reflection waveguides.
\newblock {\em Appl. Phys. Lett.},  92:101124, 2008.

\bibitem{Abolghasem2009}
P. Abolghasem, J. Han,  B. J. Bijlani, A. Arjmand and A. S. Helmy
\newblock Continuous-wave second harmonic generation in Bragg reflection waveguides.
\newblock {\em Opt. Express}, 17:9460, 2009.

\bibitem{Chen2017}
C.~Chen, E.~Y. Zhu, A.~Riazi, A.~V. Gladyshev, C.~Corbari, M.~Ibsen, P.~G. Kazansky, and L.~Qian.
\newblock Compensation-free broadband entangled photon pair sources.
\newblock {\em Opt. Express}, 25:22667, 2017.

\bibitem{Gunthner-2015}
T.~G\"{u}nthner, B.~Pressl, K.~Laiho, J.~Ge{\ss}ler, S.~H\"{o}fling, M.~Kamp, C.~Schneider, and G.~Weihs.
\newblock Broadband indistinguishability from bright parametric downconversion  in a semiconductor waveguide.
\newblock {\em J.Opt.}, 17:125201, 2015.

\bibitem{Abolghasem2009_2}
P.~Abolghasem, J.~Han, B.~J. Bijlani, A.~Arjmand, and A.~S. Helmy.
\newblock Highly efficient second-harmonic generation in monolithic matching  layer enhanced Al$_x$Ga$_{1-x}$As Bragg reflection waveguides.
\newblock {\em IEEE Photon. Technol. Lett.}, 21:1462, 2009.

\bibitem{Grice2001}
W.~P. Grice, A.~B. U'Ren, and I.~A. Walmsley.
\newblock Eliminating frequency and space-time correlations in multiphoton  states.
\newblock {\em Phys. Rev. A}, 64:063815, 2001.

\bibitem{W.P.Grice1997}
W.~P. Grice and I.~A. Walmsley.
\newblock Spectral information and distinguishability in type-II  down-conversion with broadband pump.
\newblock {\em Phys.~Rev.~A}, 56:1627, 1997.

\bibitem{Avenhaus2009}
M.~Avenhaus, M.~V. Chekhova, L.~A. Krivitsky, G.~Leuchs, and Ch. Silberhorn.
\newblock Experimental verification of high spectral entanglement for pulsed  waveguided spontaneous parametric down-conversion.
\newblock {\em Phys.~Rev.~A}, 79:043836, 2009.

\bibitem{Zhukovsky2013}
S.~V. Zhukovsky, L.~G. Helt, D.~Kang, P.~Abolghasem, A.~S. Helmy, and J.~E.  Sipe.
\newblock Analytical description of photonic waveguides with multilayer  claddings: Towards on-chip generation of entangled photons and bell states.
\newblock {\em Opt. Comm.}, 301:127, 2013.

\bibitem{Jin2018}
R.-B. Jin, R.~Shiina, and R.~Shimizu.
\newblock Quantum manipulation of biphoton spectral distributions in a 2D frequency space toward arbitrary shaping of a biphoton wave packet.
\newblock {\em Opt. Express}, 26:21153, (2018).

\bibitem{Graffitti2018}
F.~Graffitti, P.~Barrow, M.~Proietti, D.~Kundys, and A.~Fedrizzi.
\newblock Independent high-purity photons created in domain-engineered  crystals.
\newblock {\em Optica}, 5:514, 2018.

\bibitem{Chen2018}
H.~Chen, S.~Auchter, M.~Prilm\"uller, A.~Schlager, T.~Kauten, K.~Laiho, B.~Pressl, H.~Suchomel, M.~Kamp, S.~H\"ofling, C.~Schneider, and G.~Weihs.
\newblock Time-bin entangled photon pairs from Bragg-reflection waveguides.
\newblock {\em APL Photonics}, 3:080804, (2018).

\bibitem{Eckstein2008}
A.~Eckstein and C.~Silberhorn.
\newblock Broadband frequency mode entanglement in waveguided parametric downconversion.
\newblock {\em Opt. Lett.}, 33:1825, 2008.

\bibitem{Hill1997}
S. Hill and W. K. Wootters.
\newblock Entanglement of a pair of quantum bits.
\newblock {\em  Phys. Rev. Lett.}, 78:5022, 1997.


\end{thebibliography}
\end{document}